\documentclass[prl,twocolumn,superscriptaddress,amsmath,amssymb,showpacs,showkeys,shortbibliography]{revtex4-1}

\newcommand{\beq}{\begin{equation}}
\newcommand{\eeq}{\end{equation}}
\newcommand{\bea}{\begin{eqnarray}}
\newcommand{\eea}{\end{eqnarray}}

\begin{document}

\title{Period finding with adiabatic quantum computation}

\author{Itay Hen}
\email{itayhen@isi.edu}
\affiliation{Information Sciences Institute, University of Southern California, Marina del Rey, CA 90292}

\date{\today}

\pacs{03.67.Ac,03.67.Lx}

\begin{abstract}
We outline an efficient quantum-adiabatic algorithm that solves Simon's problem, in which one has to determine the `period', or xor-mask, of a given black-box function. We show that the proposed algorithm is exponentially faster than its classical counterpart and has the same complexity as the corresponding circuit-based algorithm. Together with other related studies, this result supports a conjecture that the complexity of adiabatic quantum computation is equivalent to the circuit-based computational model in a stronger sense than the well-known, proven polynomial equivalence 
between the two paradigms. We also discuss the importance of the algorithm and its theoretical and experimental implications for the existence of an adiabatic version of Shor's integer factorization algorithm that would have the same complexity as the original algorithm.
\end{abstract}

\maketitle

\section{Introduction}
Theoretical research on quantum computing is motivated by the exciting
possibility that quantum computers are inherently more efficient than
classical computers due to the advantages that the laws of quantum mechanics
provide, such as parallelism, tunneling and entanglement. 
The actual implementation of quantum computing devices is however hindered by many challenging difficulties, the most 
prominent  of which being the control or removal of quantum decoherence~\cite{schlosshauer:04}. 
Recent theoretical work~\cite{amin:09a, amin:09c, childs:01}, as well as promising experimental research findings~\cite{mcgeoch:13,berkley:13,johnson:11} in the field of Adiabatic Quantum Computing (AQC)
suggest that a leading candidate to be the first device to solve 
practical classically-hard problems using quantum principles is the so called `quantum annealer', 
which implements the simple yet potentially-powerful quantum-adiabatic algorithmic approach
proposed by Farhi {\it et al.}~\cite{farhi_long:01} about a decade ago.

Within the Quantum Adiabatic Algorithm (QAA) approach, the solution to an optimization problem is encoded in the ground state of a `problem' Hamiltonian
$\hat{H}_p$. To find the solution, the system is prepared in the ground state of another `driver' Hamiltonian
$\hat{H}_d$ that must not commute $\hat{H}_p$ and has a ground state that is fairly easy to prepare. 
The Hamiltonian of the system is then slowly modified from $\hat{H}_d$ to
$\hat{H}_p$, using the linear interpolation, i.e., $\hat{H}(s)=s \hat{H}_p +(1-s) \hat{H}_d$
where $s(t)$ is a parameter varying smoothly with time from 0 to 1. 
If this process is done slowly enough, the system will stay close to the ground state of the
instantaneous Hamiltonian throughout the evolution~\cite{kato:51,messiah:62}, so that one finally
obtains a state close to the ground state of $\hat{H}_p$.  At this point,
measuring the state will give the solution of the original problem with high
probability.  The running time $\mathcal{T}$ of the algorithm determines the
efficiency, or complexity, of the QAA and normally scales inversely with a small power of the minimum gap of the system~\cite{farhi_long:01,jansen:07}.

The experimental studies mentioned above, as well as other theoretical work such as the theorem of polynomial
equivalence between AQC and circuit-based computing~\cite{aharonov:07,mizel:07}, provide ample motivation
for searching for adiabatic algorithms that would reveal the inherent potential encompassed in AQC.
However, despite intensive research (see, e.g., Ref.~\cite{hogg:03,farhi:02,farhi:08,altshuler:09,young:10,hen:11,hen:12} and references therein), to date, most attempts aimed at finding practical efficient AQC algorithms have failed.

Valuable insight into why this is so, may be obtained by considering, in the context of AQC, problems for 
which there are known circuit-based algorithms that exhibit `quantum speedups'. 
Devising `optimal' adiabatic algorithms for these, that would have the same complexity as their circuit analogues, may reveal important features of AQC that have perhaps not been properly utilized thus 
far~\footnote{In that respect, the prescription proposed by the polynomial equivalence theorem~\cite{aharonov:07} may be 
viewed as `artificial' by construction and impractical for actual implementation.  Additionally, it will in general be suboptimal, 
as it is expected to yield unnecessary polynomial overheads.}.
Several known examples for adiabatic analogues to circuit-based algorithms are
Grover's unstructured search problem~\cite{roland:02} and Quantum Counting~\cite{hen:13a},
both of which have successfully recovered the quadratic quantum speedups exhibited by their
circuit-model counterparts, thereby hinting at the potential power of AQC.

Perhaps the most important class of problems exhibiting quantum speedups is that of the Hidden Subgroup Problem 
(HSP)~\cite{lomont:04}, which 
encompasses most of the quantum algorithms found
so far that are \emph{exponentially} faster than their classical counterparts (e.g., period-finding, order-finding and discrete logarithms). 
In HSPs, one is given a function from a finitely generated group to a finite set. The function is known to be constant on the 
cosets of a subgroup whose generating set needs to be found. 
Finding adiabatic algorithms for problems in this class that have the same complexities as their circuit-based counterparts, is therefore of considerable importance. 

The simplest HSP considered in the context of AQC is the problem of Deutsch and Josza~\cite{deutsch:92},
which has been shown to have an $O(1)$ complexity, similarly to the circuit-based result, and is therefore exponentially faster than the 
deterministically classical algorithm~\cite{hen:14} (see also Ref.~\cite{sarandy:05,panduranga:03}). 
Here, we devise a QAA for a more challenging HSP, namely, Simon's problem~\cite{simon:94}, which was the first to exhibit exponential speedup
over classical probabilistic algorithms. 

In Simon's problem one is asked to find the `period' of a function from $n$-bit strings to $(n-1)$-bit strings that is `promised' to be constant only over inputs that are related by an unknown xor mask (the period).
Although of a somewhat artificial nature, the significance of Simon's problem stems from the fact that it shares many similarities with, and in fact was also the motivation for, Shor's integer factorization algorithm~\cite{shor:94}, which in itself is based on a `genuine' period-finding problem. 
A derivation of a quantum adiabatic integer factorization algorithm that would have the same complexity as Shor's original circuit, together with the aforementioned significant experimental advancements in the construction and operation of quantum annealers, may have considerable theoretical as well as practical impact and implications in the field of Quantum Computing and beyond it, as Shor's algorithm may be used to break the RSA public-key cryptosystem. In that sense, a quantum-adiabatic Simon's problem algorithm is an important first step towards that goal. 

Before we consider the full complexity of Simon's problem, let us first study the related, albeit simpler, HSP
devised by Bernstein and Vazirani (BV)~\cite{bernstein:97}.
The algorithm given here will help lay the groundwork for the discussion of the more elaborate algorithm for Simon's problem. 

\section{The Bernstein-Vazirani problem}

In the BV problem, one is given a black box that evaluates the function 
\beq \label{eq:bv}
f(w)=  \left( \sum_{k=0}^{n-1} w_k a_k \right) \mod \, 2\,,
\eeq
where $w_k$ and $a_k$ \hbox{($k=0..n-1$)} are the bits of the two integers $w$ and $a$, respectively,
and the function $f(\cdot)$ takes the integer $w$ into the modulo-2 sum of the products of
corresponding bits of $a$ and $w$. The task is to find the integer $a$ with as few queries of $f(\cdot)$ as possible. 

While the best classical algorithm to do this requires $n$ queries of $f(\cdot)$ [e.g., with the inputs \hbox{$w=2^k$} \hbox{($k=0..n-1$)} revealing the bits of $a$ one at a time], quantum circuit-based algorithms have been shown to require $O(1)$ queries to perform the same task~\cite{bernstein:97}. Here, we show that, similarly to the quantum circuit algorithm, there is a corresponding adiabatic algorithm that requires a running time that does not scale with input size. 

To show this, we first note that the function $f(\cdot)$ can be encoded as the following problem Hamiltonian:
\bea \label{eq:hpbv}
\hat{H}_p&=&
- \frac1{2} \sum_{w \in \{0,1\}^{n}} \left( |w\rangle \langle w |\right)^{(A)} \otimes \left[ 1 + (-1)^{f(w)} \sigma_z\right]^{(B)}
\nonumber\\
&=& - \sum_{w \in \{0,1\}^{n}} \left( |w\rangle \langle w |\right)^{(A)} \otimes  \left( |f(w)\rangle \langle f(w) |\right)^{(B)}\,.
\eea
Here, $\sigma_z$ ($\sigma_x$) denotes a Pauli matrix in the $z$-direction ($x$-direction) , the letter $(A)$ labels the $n$-bit subspace for the input states $|w\rangle$ and $(B)$ denotes a $1$-bit `output' subspace for the outcome 
\hbox{$|f(w)\rangle=|0\rangle$} 
or $|1\rangle$ (the state $|0\rangle$ shall represent a spin pointing in the positive $z$-direction).
For all configurations $|w\rangle$ of the input bits for which $f(w) = 1$, $\hat{H}_p$ has value $-1$ if the output qubit is $|1\rangle$ and $0$ if the output qubit is $|0\rangle$.
For all configurations $|w\rangle$ for which $f(w) = 0$, $\hat{H}_p$ has value $-1$ if the output qubit is $|0\rangle$ and $0$ otherwise.
Combining both cases, we see that the ground state of $\hat{H}_p$ has energy $-1$ and is $2^n$-fold degenerate: 
\hbox{$|w\rangle^{(A)} \otimes |f(w)\rangle^{(B)}$} for all $w \in \{0,1\}^{n}$. 
The action of $\hat{H}_p$ on `classical' configurations $|w\rangle \otimes |0\rangle$ will give an eigen-energy of $-1$ if $f(w)=0$ 
and zero otherwise. 
The above encoding may therefore be considered as the hermitian analogue of the unitary oracle of the analogous circuit-based problem. 
It should be noted nonetheless, that as an oracle, the problem Hamiltonian for the Bernstein-Vazirani problem suggested above (as well as that of Simon's problem discussed next) is not expected to, nor will it be in the general case, easily implementable. This is of course also true for the corresponding classical oracle as well as the unitary oracle of the quantum circuit-model algorithm.
 
Before moving on to describing a quantum adiabatic algorithm for the BV problem, let us first make the following simple yet important observation. 
Given a system described by a Hamiltonian
of the form \hbox{$\hat{H} = \sum_w \left(|w\rangle \langle w |\right)^{(A)} \otimes \hat{H}^{(B)}_w$} where $w$ labels basis states in subsystem $(A)$, and an initial state 
\hbox{$|\psi (t=0)\rangle=\sum_w c_w |w\rangle^{(A)} \otimes |\psi(t=0)\rangle_w^{(B)}$} where $c_w$ are complex coefficients obeying 
\hbox{$\sum_w |c_w|^2=1$}, the state of the system at time $t$ may be written as \hbox{$|\psi (t)\rangle=\sum_w c_w |w\rangle^{(A)} \otimes |\psi(t)\rangle_w^{(B)}$} where \hbox{$|\psi(t)\rangle_w^{(B)}$} is a wave-function evolving under $\hat{H}^{(B)}_w$. 
For each state $|w\rangle$ in subsystem $A$, there is an independently evolving state in subsystem $B$ that is not affected by the evolutions 
corresponding to other states $|w'\rangle$. This statement may be readily verified by plugging $|\psi (t)\rangle$ into the Schr\"odinger equation for $\hat{H}$. The above `parallelism' will be used in what follows to construct many-qubit adiabatic Hamiltonians 
that would simultaneously execute a multitude of two-level adiabatically-evolving systems. 
Since the runtime for the evolution of each subsystem will not depend on the size of the entire system, the resultant total evolution 
will likewise not require a running time that scales with system size. This unusual property~\cite{farhi:02,young:10,hen:12} will be key in the construction of the adiabatic BV and Simon's algorithms, which we address next.

To construct a QAA for the BV problem we shall choose the driver Hamiltonian:
\bea \label{eq:hdbv}
\hat{H}_d &=&\frac1{2} \left[  1^{(A)} \otimes \left( 1 - \sigma_x\right)^{(B)} \right] \\\nonumber
&=& \frac1{2} \sum_{w \in \{0,1\}^{n}} \left( |w\rangle \langle w|\right)^{(A)} \otimes  \left( 1 - \sigma_x\right)^{(B)} \,.
\eea
[We shall henceforth omit the subspace labels $(A)$ and $(B)$ in favor of a more compact notation.]
The total Hamiltonian can therefore be written as 
\beq \label{eq:H}
\hat{H} = \sum_{w \in \{0,1\}^{n}} |w\rangle \langle w | \otimes \hat{H}_w \,,
\eeq
where each $\hat{H}_w$ acts only inside the second subspace and, following Eqs.~(\ref{eq:hpbv}),~(\ref{eq:hdbv}) and ~(\ref{eq:H}), can be written as 
\beq \label{eq:Hsub}
\hat{H}_w = \frac1{2} \left[ (1-s)   \left( 1 - \sigma_x\right) - s  \left( 1 + (-1)^{f(w)} \sigma_z\right) \right]\,.
\eeq

Having set up the problem and driver Hamiltonians, the QAA is as follows. 

\begin{itemize}
\item
Step 1: 
As the initial state, prepare the equal superposition of all computational-basis states 
\beq
|\psi(t=0)\rangle=\frac1{\sqrt{2^n}} \sum_{w \in \{0,1\}^{n}} |w\rangle \otimes |+\rangle = |+\rangle \otimes |+\rangle \,,
\eeq 
where $|+\rangle$ ($|-\rangle$) denotes the state of all spins of the appropriate subsystem pointing in the positive (negative) $x$-direction
(note that this is easily done).
\item 
Step 2: Evolve the system under the Hamiltonian, Eq.~(\ref{eq:H}), where $s$, the adiabatic parameter, is varied at a constant rate
from $s=0$ at $t=0$ to $s=1$ at $t=\mathcal{T}$ for some $n$-independent runtime $\mathcal{T}$.

Utilizing the observation made above, the evolution of the total wave-function is split to $2^n$ decoupled $1$-qubit evolutions, 
each described by the simple gapped two-level system $\hat{H}_w$, Eq.(\ref{eq:Hsub}), with the initial condition 
\hbox{$|\psi(t=0)\rangle_w = |+\rangle$}. Therefore, an appropriately-long (but $n$-independent) runtime $\mathcal{T}$ will evolve each
\hbox{$|\psi(t=0)\rangle_w$} into the ground state of the problem Hamiltonian of $\hat{H}_w$, namely, into 
\hbox{$|\psi(t=\mathcal{T})\rangle_w=|f(w)\rangle$}
with an arbitrarily high accuracy that depends on $\mathcal{T}$. It is important to note that there will be no relative phases 
between the different final states $|f(w)\rangle$. As a result, the final state of the entire system
will be 
\beq
|\psi(t=\mathcal{T})\rangle = \frac1{\sqrt{2^n}} \sum_{w \in \{0,1\}^{n}} |w\rangle \otimes |f(w)\rangle \,.
\eeq 
The above decoupled evolution may be viewed as `adiabatic quantum parallelism' as it describes the (coherent) evolution
of many adiabatic systems that evolve in parallel, such that the coherent superposition of the end states may be utilized 
to extract important information about the system (as is shown in the next step).
\item
Step 3: 
At the end of the evolution, perform an $x$-basis measurement of the qubit of the second subsystem. Noting that $|f(w)\rangle$ may be written as:
\beq
|f(w)\rangle = \frac1{\sqrt{2}} \left( |+\rangle + (-1)^{f(w)}|-\rangle \right) \,, 
\eeq
the measurement will collapse the wave-function into one of
two states with equal probability:
\bea
|\psi_{+} \rangle &=& \frac1{\sqrt{2^n}} \sum_{w \in \{0,1\}^{n}} |w\rangle \otimes |+\rangle=|+\rangle \otimes |+\rangle\\
\nonumber \textrm{or}
\\
|\psi_{-} \rangle &=& \frac1{\sqrt{2^{n}}} \sum_{w \in \{0,1\}^{n}} (-1)^{f(w)}|w\rangle  \otimes | - \rangle \\\nonumber
&=&\bigotimes_{k=0}^{n-1} \Big(|0\rangle + (-1)^{a_k}|1\rangle \Big)  \otimes | - \rangle \,.
\eea
Now, if $|\psi_+\rangle$ is obtained, then one must go back to Step 1 and restart the algorithm, as the resulting state will contain no information about $a$ (in fact, $|\psi_+\rangle$ is the initial state). Otherwise, the resulting state will encode all the bits of $a$, namely $a_k$, each corresponding to a different orientation of the spin along the $x$-axis. 
The probability that the algorithm would fail after $r$ runs of the QAA is exponentially small, namely, $P_{\textrm{failure}}=2^{-r}$ but more importantly, 
$n$-independent. 
\end{itemize}

We have thus shown that the BV problem can be solved using a quantum adiabatic algorithm 
in $O(1)$ runtime, i.e., with complexity that is equivalent to that of the corresponding circuit-based algorithm, which requires
only one call to the unitary quantum oracle.
 
Before moving on, we note here that while the problem Hamiltonian, Eq.~(\ref{eq:hpbv}), is diagonal, the state of the system at the end of the adiabatic evolution is a superposition of computational-basis states. In practice, this would require a high level of coherence, a property that is sometimes not strictly necessary for other adiabatic quantum computations. 

We now turn to address the somewhat more complicated HSP devised by Simon~\cite{simon:94}, utilizing for that purpose 
the concepts presented above. 

\section{\label{sec:sop}Simon's problem}
The problem devised by Simon~\cite{simon:94} was the first to demonstrate a quantum algorithm solving a black-box problem 
that is exponentially faster than any probabilistic classical algorithm.
In Simon's problem, we are given a promise that there is an $n$-bit positive integer $a$ such that for any two $n$-bit inputs $w\neq y$, 
a black-box function $g(\cdot): \{0,1\}^n \to \{0,1\}^{n-1}$ would output the $(n-1)$-bit integers $g(w)=g(y)$ if and only if $w \oplus y = a$. The symbol $\oplus$ denotes
here  the bitwise-xor operation. We are then asked to find the `period' (or, `xor-mask') of the function, namely $a$,
with as few queries of $g(\cdot)$ as possible. While classical algorithms require $O(2^{n/2})$ queries of $g(\cdot)$, Simon~\cite{simon:94} devised a circuit-based quantum algorithm that can perform this task with $O(n)$ queries~\cite{mermin:07}. 

In the context of QAA, the oracle $g(\cdot)$ can be encoded as the problem Hamiltonian
\beq
\hat{H}_p = \sum_{w \in \{0,1\}^n} |w\rangle \langle w| \otimes \sum_{y \in \{0,1\}^{n-1}} h[y, g(w)] |y \rangle \langle y | \,,
\eeq
where $h[y, g(w)]$ is the Hamming distance between $y$ and $g(w)$. 
Here, $y \in \{0,1\}^{n-1}$ labels a bit configuration in the second subsystem. 
The ground state of the total Hamiltonian is $2^n$-fold degenerate, namely, $|w\rangle \otimes |g(w)\rangle$ for all \hbox{$w \in \{0,1\}^{n}$}.
An alternative and more illuminating way to write $\hat{H}_p$ would be
\beq
\hat{H}_p=\frac1{2}  \sum_{w \in \{0,1\}^n} |w\rangle \langle w| \otimes \sum_{k=1}^{n-1} \left[1-(-1)^{g_k(w)} \sigma_z^k \right]
\eeq
where $\sigma_z^k$ denotes a Pauli matrix acting on the $k$-th bit of the second subsystem and $g_k(w)$ is the $k$-th bit of $g(w)$.
This illustrates the fact that the spins of the second subsystem for each $|w\rangle$ are not coupled
to one another. 

Similarly to the BV problem, we shall choose the driver Hamiltonian for Simon's algorithm to be 
the transverse-field Hamiltonian on the second subsystem, namely: 
\beq
\hat{H}_d = \frac1{2} \left[ 1 \otimes \sum_{k=1}^{n-1} (1-\sigma^k_x)\right]\,.
\eeq
The total Hamiltonian can therefore be written as \hbox{$\hat{H} = \sum_w |w\rangle \langle w | \otimes \hat{H}_w$}, where 
$\hat{H}_w$ acts only on the second subsystem and can be written as 
\beq
\hat{H}_w = \frac1{2} \sum_{k=1}^{n-1} \left[ (1-s)   (1-\sigma^k_x)+ s  \left( 1-(-1)^{g_k(w)} \sigma^k_z \right) \right] \,.
\eeq
Having constructed the problem and driver Hamiltonians, the algorithm is as follows. 
\begin{itemize}
\item
Step 1: 
As the initial state, prepare the equal superposition of all computational-basis states:
\beq
|\psi(t=0)\rangle = \frac1{\sqrt{2^n}} \sum_{w \in \{0,1\}^{n}} |w\rangle \otimes |+\rangle = |+\rangle \otimes |+\rangle \,. 
\eeq 
\item 
Step 2: Evolve the system using a constant-rate adiabatic evolution for a fixed (i.e., an $n$-independent) amount of time $\mathcal{T}$. 
The Hamiltonians $\hat{H}_w$ are each an uncoupled system of $n$-spins. 
It should therefore be clear that the wave-functions of the $n$-spins will evolve simultaneously and independently into the ground state of the problem Hamiltonian for every $w$ (for an appropriately-chosen $\mathcal{T}$), yielding
\hbox{$|\psi(t=\mathcal{T})\rangle_w=|g(w)\rangle$}.  The final state of the entire system will therefore be 
\bea
|\psi(t=\mathcal{T})\rangle= \frac1{\sqrt{2^n}} \sum_{w \in \{0,1\}^{n}} |w\rangle \otimes |g(w)\rangle
\\\nonumber = \frac1{\sqrt{2^{n-1}}} \sum_{w \in \{0,1\}^{n}/g} (|w\rangle + |w \oplus a \rangle) \otimes |g(w)\rangle \,.
\eea  
\item
Step 3: 
A $z$-basis measurement of the second subsystem will collapse the wave-function into a state 
\bea
|\psi_{w_*}\rangle = \frac1{\sqrt{2}} (|w_*\rangle + |w_* \oplus a \rangle) \otimes |g(w_*)\rangle \nonumber \\
= \frac1{\sqrt{2^{n+1}}} \sum_{x \in \{0,1\}^{n}}  (-1)^{w_* \cdot x} \left[ 1+(-1)^{f(x)}\right] |x\rangle \otimes |g(w_*)\rangle \nonumber\\
= \frac1{\sqrt{2^{n-1}}}\sum_{f(x) = 0}  (-1)^{w_* \cdot x} |x\rangle \otimes |g(w_*)\rangle \,,\nonumber\\
\eea
where $w_*$ is a random bit configuration, $f(\cdot)$ is the BV function, Eq.~(\ref{eq:bv}), and $|x\rangle$ are $x$-basis states.
A subsequent $x$-basis measurement of the first subsystem will, similarly to the circuit-based algorithm, yield
a state $|x_*\rangle$ whose bit representation is orthogonal to that of $a$ (modulo 2), namely, obeying $f(x)=0$.
\item
Step 4: As in the circuit-based model, each time Steps 1 through 3 are repeated, with high probability, a different state $|x\rangle$ that is orthogonal to $a$ will be obtained~\cite{mermin:07}. The different runs will thus produce $(n-1)$ linear equations in the bits of $a$ that are highly likely 
to be independent of one another~\footnote{There is an exponentially small probability that the $(n-1)$ equations will be linearly dependent.
In this case, more runs will be needed. Altogether, $n+20$ runs will normally be sufficient~\cite{mermin:07}.}. 
Solving the resultant system of equations will produce the bits of $a$. The total query complexity of the entire process is therefore $O(n)$, 
similarly to the circuit-based case.
\end{itemize}

\section{\label{sec:conc}Summary and conclusions}
We have outlined two quantum adiabatic algorithms to solve the HSPs proposed by Bernstein and Vazirani~\cite{bernstein:97}
and the `period-finding' problem devised by Simon~\cite{simon:94}. For the BV problem, an algorithm with an $O(1)$ complexity has been found [in contrast to its $O(n)$ classical counterpart], whereas for Simon's problem we have devised an efficient $O(n)$ algorithm which is to be 
contrasted with the exponentially-slow classical algorithm. The complexities of both algorithms have been found to be the same
as the analogous circuit-based results. 

The importance of the algorithm constructions given above, lies in the fact that these do not use the (suboptimal) 
polynomial-equivalence prescription of `translating' circuit-based algorithms to quantum-adiabatic ones~\cite{aharonov:07}.  The adiabatic Simon's algorithm given here
is therefore one of very few `genuine' quantum-adiabatic algorithms that provide an exponential speedup over probabilistic classical algorithms,
and with no complexity overhead, thereby exhibiting the full potential power encompassed in AQC. 

In addition, along with other recent results comparing circuit-based and adiabatic algorithms~\cite{hen:13a,hen:14}, 
this study provides further corroboration 
to a conjecture that suitably-constructed quantum adiabatic algorithms have the same complexities as their circuit-based counterparts.
This claim is a stronger statement than the theorem establishing a (weaker) polynomial equivalence~\cite{aharonov:07, mizel:07} between the two computing paradigms, and still awaits formal proof (or disproof).  

We note here that the algorithms proposed here are not of the `traditional' AQC form, in the sense that the ground-states of the adiabatic Hamiltonians presented here are highly degenerate. As was also briefly discussed above, it follows that the algorithms suggested here
may therefore require higher levels of coherence than that needed for traditional AQC. An important question that still remains is whether or not one could construct adiabatic algorithms that are of the same complexity as their circuit-model counterparts, but that would also possess the robustness of the `usual' AQC. This is particularly of
significance if one is to consider the actual implementation of algorithms similar to the ones presented here, 
on real, finite-temperature, noisy quantum annealers.
 
Finally, it is plausible to believe that the algorithm given above to Simon's problem will provide the basis, as well as further motivation,
for devising a quantum adiabatic algorithm for integer factorization, as the two problems are
known to share many similarities. Such an algorithm would presumably have the same complexity as 
Shor's original algorithm~\cite{shor:94} and will therefore be more efficient than the analogous polynomial-equivalence construction.  An adiabatic Shor's algorithm would require replacing Simon's black-box function with the encoding of a `real' periodic function and, as a next step, would necessitate the construction of an adiabatic algorithm for the Quantum Fourier Transform.  
An efficient quantum-adiabatic integer factorization algorithm that will be implementable on a real many-qubit
quantum annealer in the near future, would undoubtedly have considerable theoretical as well as practical implications both in the field of Quantum Computing and well beyond it.

\acknowledgments
We thank Peter Young, Siddharth M. Krishnan, David Gosset and Mohammad Amin for useful comments and discussions. 

\bibliographystyle{ieeetr}
\bibliography{refs}{}

\end{document}